\documentclass{iau307}
\usepackage{graphicx}
\usepackage{natbib}
\usepackage{url}
\usepackage{dtklogos}
\bibpunct{(}{)}{;}{a}{}{,}

\title[Interferometric Snapshot Survey] 
{AMBER/VLTI Snapshot Survey on 
Circumstellar Environments}

\author[Th. Rivinius et al.] 
       {Th.~Rivinius$^{1}$ \and W.J.~de Wit$^{1}$ \and Z.~Demers$^{2,3}$ \and
         A.~Quirrenbach$^{3}$ \and the VLTI Science Operations Team}

\affiliation{$^{1}$ESO - European Organisation for Astronomical
  Research in the Southern Hemisphere, Chile \\ email: {\tt triviniu@eso.org} \\[\affilskip]
$^{2}$Department of Physics and Astronomy, The University of Western Ontario, London, Canada\\[\affilskip]
$^{3}$Landessternwarte K\"onigstuhl, D-69117, Heidelberg, Germany
}

\pubyear{2014}
\volume{307} 
\pagerange{}
\jname{New windows on massive stars: asteroseismology, interferometry, and spectropolarimetry}
\editors{G. Meynet, C. Georgy, J.H. Groh \& Ph. Stee, eds.}

\begin{document}

\maketitle

\begin{abstract}
OHANA is an interferometric snapshot survey of the gaseous circumstellar
environments of hot stars, carried out by the VLTI group at the Paranal
observatory. It aims to characterize the mass-loss dynamics (winds/disks) at
unexplored spatial scales for many stars. The survey employs the unique
combination of AMBER's high spectral resolution with the unmatched spatial
resolution provided by the VLTI. Because of the spatially unresolved central
OBA-type star, with roughly neutral colour terms, their gaseous environments
are among the easiest objects to be observed with AMBER, yet the extent and
kinematics of the line emission regions are of high astrophysical
interest. 
\keywords{stars: winds, outflows, circumstellar matter}
\end{abstract}

\firstsection 
\section{Introduction}
The {\bf O}bservatory survey at {\bf H}igh {\bf AN}gular resolution of {\bf
  A}ctive OB stars (OHANA) combines high spectral with high spatial resolution
across the Br$\gamma$ and He{\sc i}$\lambda$2.056 lines to characterize the
dynamics of winds and disks. It is carried out by the VLTI group at the
Paranal observatory with the three-beam combining instrument AMBER
\citep{2007A&A...464....1P}. The survey was designed to make use of the
observing time not requested by other programs, usually due to bad weather or
unsuitable local sidereal time slots.

\section{Observations and Data Reduction}
The survey targets consist of twelve bright Be stars, thirteen O and B type
supergiants, and one interacting binary (see {Table~1}). Almost 300
observations were obtained. By design, namely targeting quantities relative
  to the adjacent continuum, no calibrators were observed. However, in some
  nights calibrators, taken for technical purpose or other programs using the
  same setup, are available. These have been added to the database.

Basic data reduction was performed with amdlib, v3.0.6
\citep{2007A&A...464...29T,2009A&A...502..705C}, and then processed further
with idl. In particular:

\begin{itemize}
\item The pixel shifts between the spectral channels were a matter of concern,
  and seem not to be entirely stable. Whether this is a real effect or a
  consequence of noise affecting the determination of the shift is under
  investigation.
\item Since the program aims for relative quantities, which do not suffer from
  degradation of absolute visibility, 100\% of the frames were selected for
  display in this work. However, the final reduction includes several lower
  selection ratios as well.
\item In the case of continuous observations of more than 30 minutes, ($u,v$)
  points were merged into 30 minute bins.
\item Intensity spectra were extracted, and the absolute wavelength scale
  corrected using telluric lines. The flux continuum was normalized to unity.
\item Visibilities in the continuum were normalized to unity, phases in the
  continuum to zero. If calibrators were taken, these were used to check for
  and eventually remove instrumental ripples.
\item RMS in the continuum was measured for each quantity to estimate data
  quality.
\end{itemize}
The raw data have become public immediately, and the results of the final
reduction of the Br$\gamma$ observations will be made public as soon as they
are complete. The reduction of the He{\sc i}$\lambda$2.056 observations is
pending.  

\begin{table}
\begin{center}
\caption{Observed targets, spectral types, and data
   obtained. For each spectral line, the number of observations on the {\bf
     s}mall, {\bf i}ntermediate, and {\bf l}arge telescope configurations
   (s--i--l) are given.}
\label{Rivinius_OHANA_tab1}
\begin{tabular}{llccllcc}
Target             & Sp.~type   & Br$\gamma$ & He{\sc i}$\lambda$2.056&
Target             & Sp.~type   & Br$\gamma$ & He{\sc i}$\lambda$2.056\\
                   &            &  s--i--l   & s--i--l &
                   &            &  s--i--l   & s--i--l \\
\hline\\[-2ex]
\multicolumn{4}{c}{\bf Be Stars}\smallskip & \multicolumn{4}{c}{\bf OBA Supergiants} \\
$\mu$\,Cen         & B2\,Vnpe & 2--5--3 & 0--1--0 &
$\eta$\,Car        & LBV      & 22--16--5 & 3--1--0\\
$\chi$\,Oph        & B2\,Vne  & 0--0--1 & 0--0--0 &
HR\,Car            & LBV      & 4--4--1 & 2--0--0\\
$\zeta$\,Tau       & B2\,IVe-sh  & 2--1--0 & 1--0--0&
$\zeta$\,Pup       & O4\,If   & 4--6--1 & 1--2--2\\
$\delta$\,Cen      & B2\,IVne & 3--5--2 & 1--1--1&
$\iota$\,Ori       & O9\,III  & 1--2--0 & 1--0--0\\
$\epsilon$\,Cap    & B3\,Ve-sh& 1--5--4 & 0--2--0&
$\zeta$\,Ori       & O9.7\,Iab& 2--1--2 & 1--0--1\\
$\beta^1$\,Mon\,A  & B3\,Ve   & 6--8--0 & 2--1--0&
$\epsilon$\,Ori    & B0\,Iab  & 1--1--0 & 0--0--0\\
$\beta^1$\,Mon\,B  & B3\,ne   & 2--1--0 & 0--0--0&
$\kappa$\,Ori      & B0\,Iab  & 3--2--0 & 2--0--0\\
$\beta^1$\,Mon\,C  & B3\,e    & 2--1--0 & 0--0--0&
$\zeta^1$\,Sco     & B0.5\,Ia+& 0--1--3 & 0--0--1\\
P\,Car             & B4\,Vne  & 6--5--2 & 2--3--1&
$\gamma$\,Ara      & B1\,Ib   & 0--1--0 & 0--1--0\\
$\beta$\,Psc       & B6\,Ve   & 1--4--4 & 0--2--0&
HR\,6142           & B1\,Ia   & 0--0--1 & 0--0--1\\
$\eta$\,Tau        & B7\,IIIe & 0--0--0 & 1--0--0&
$\epsilon$\,CMa    & B2\,Iab  & 4--3--1 & 1--1--1\\
Electra            & B8\,IIIe & 0--0--0 & 1--0--0&
HD\,53\,138        & B3\,Ia   & 11--16--3 & 2--5--1\\[.5ex]
\multicolumn{4}{c}{\bf Interacting Binary}\smallskip & 
V533\,Car          & A6\,Iae  & 3--5--2 & 1--1--1\\
SS\,Lep            &  A1\,V\,+\,M6\,II    & 7--1--2 & 1--1--2 &
&&&\\
\hline\\[-2ex]
\end{tabular}\smallskip
\end{center}
\end{table}

\section{Data Description and First Impressions}
\begin{figure}[t]
\begin{center}
\parbox{0.5\textwidth}{\centerline{$\zeta^1$\,Sco (B0.5\,Ia+)}}%
\parbox{0.5\textwidth}{\centerline{HD\,53\,138 (B3\,Ia)}}%

\includegraphics[width=0.5\textwidth]{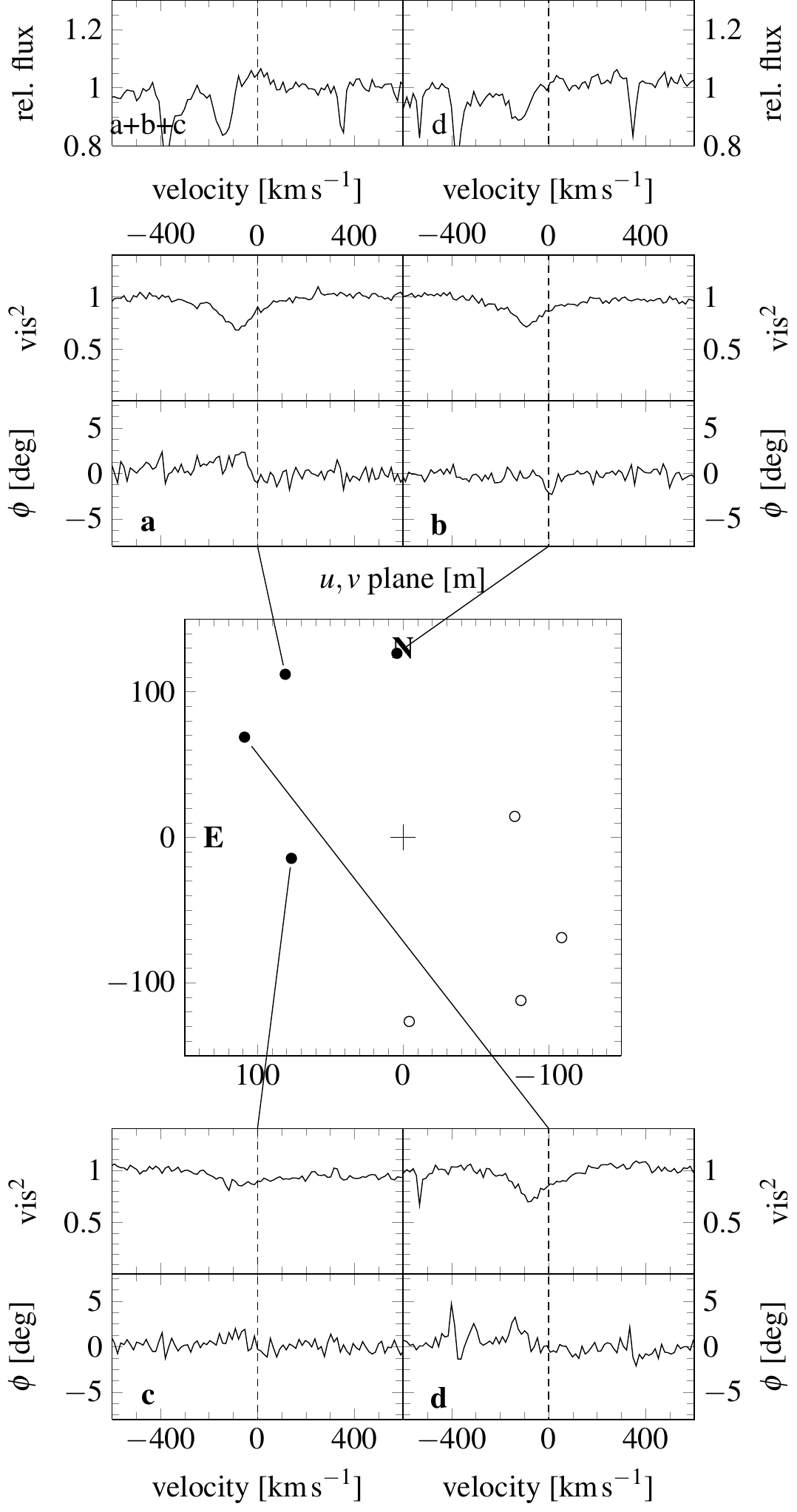}%
\includegraphics[width=0.5\textwidth]{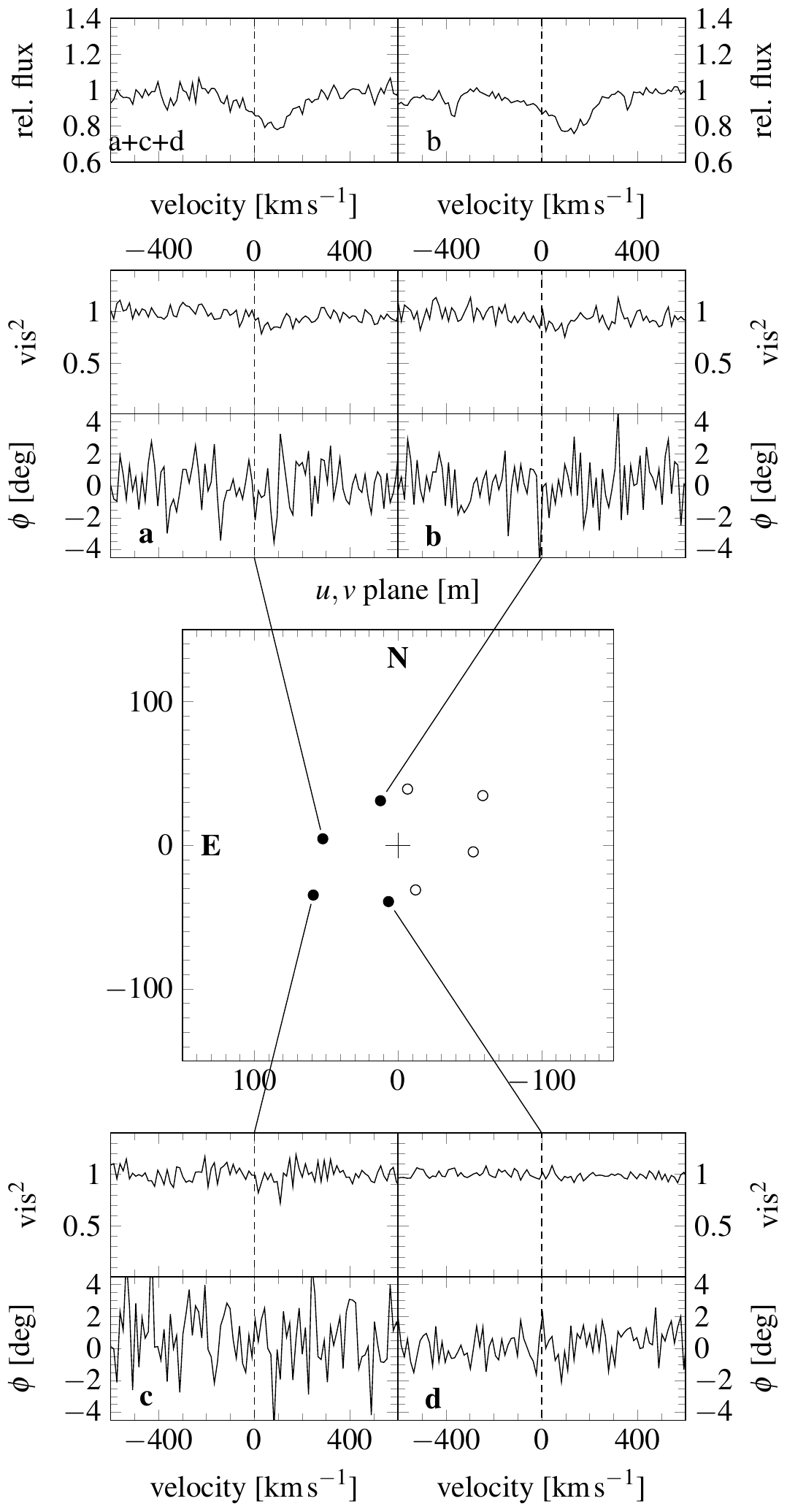}%
\caption{Example of OHANA data for the B-type supergiants $\zeta^1$\,Sco and HD\,53\,138.}
\label{Rivinius_OHANA_fig2}
\end{center}
\end{figure}

Due to the snapshot/backup/filler nature of the program, the data quality is
inhomogeneous. Typical values for a good data set are an uncertainty of the
visibility (normalized to unity) of about $\pm0.05$, and of the phase $\pm
2^\circ$, at a SNR of the combined spectrum of above 100.
Selected data sets of the target stars are shown in
Figs.~\ref{Rivinius_OHANA_fig1} and \ref{Rivinius_OHANA_fig2}. For each of the
four targets, four baselines are shown, taken from two observations. The
uppermost panels for each target show the flux spectra, then subpanels a-d show
visibility and phase (upper and lower resp.\ profiles), while the centered
panel show the ($u,v$) plane covered by the four baselines shown.

\subsection{Be Stars}

\begin{figure}[t]
\begin{center}
\parbox{0.5\textwidth}{\centerline{$\mu$\,Cen (B2\,Vnpe)}}%
\parbox{0.5\textwidth}{\centerline{$\beta^1$\,Mon (B3\,Ve)}}%

\includegraphics[width=0.5\textwidth]{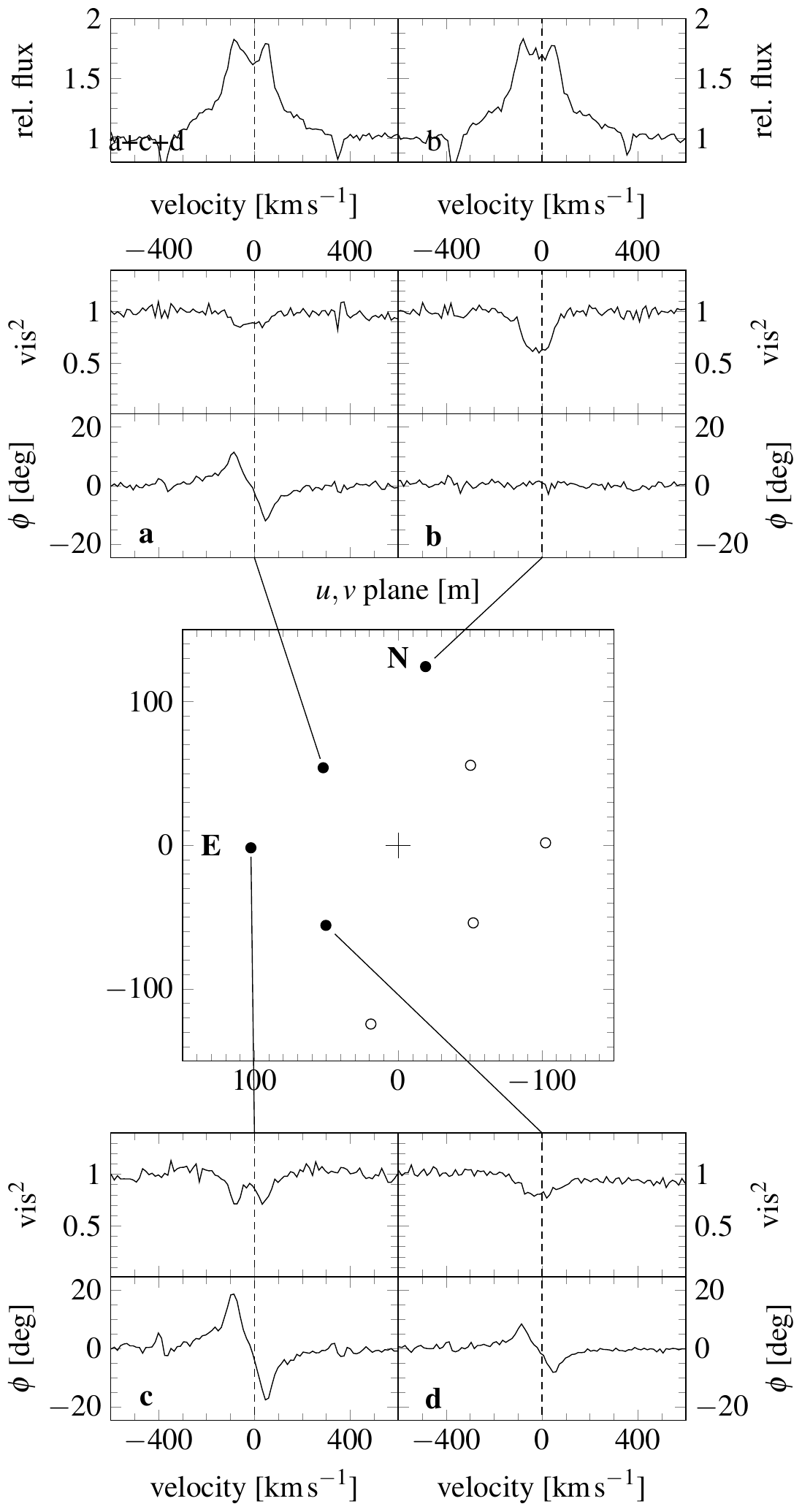}%
\includegraphics[width=0.5\textwidth]{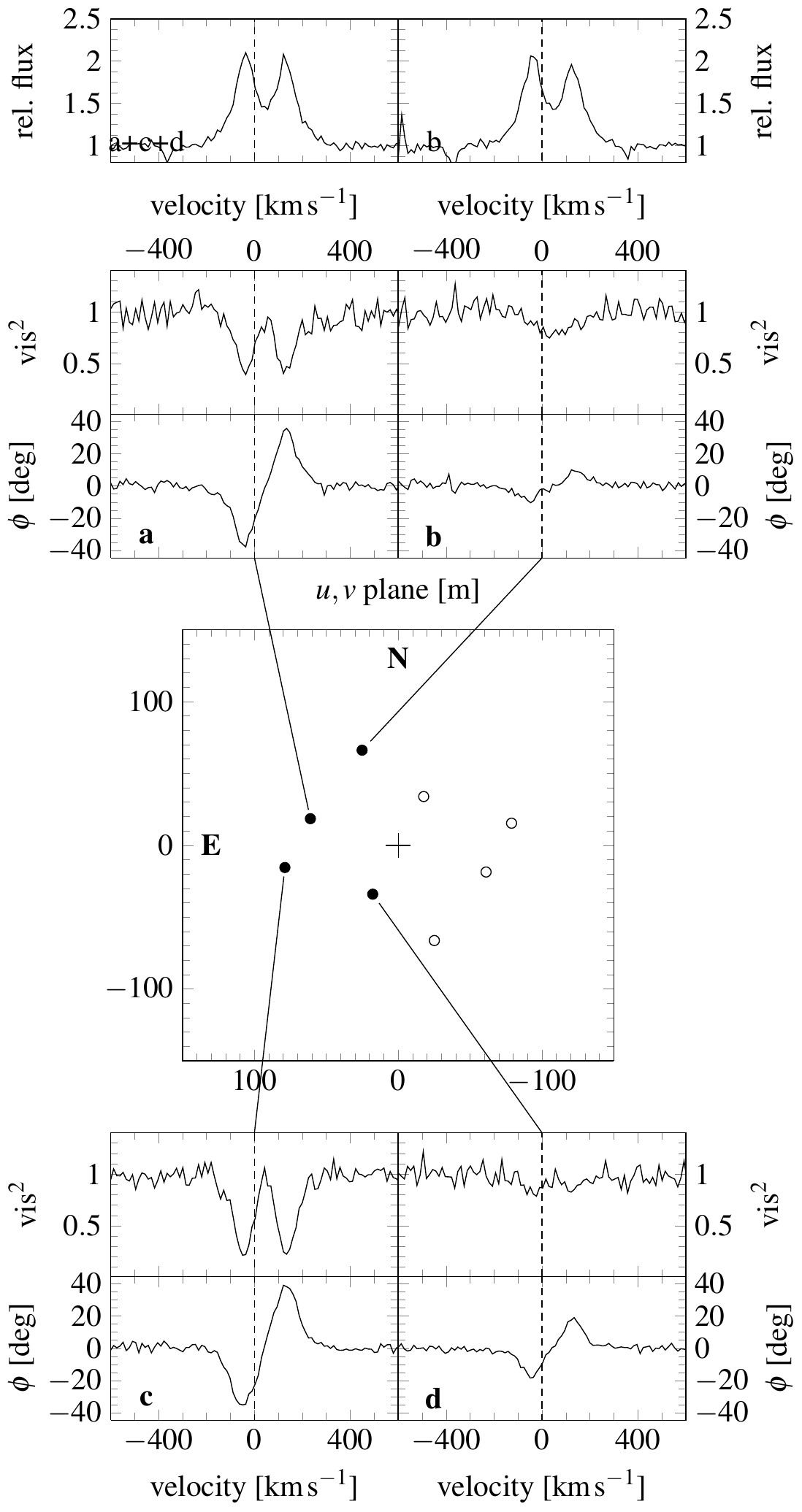}%
\caption{Example of OHANA data for the Be stars $\mu$\,Cen and $\beta^1$\,Mon.}
\label{Rivinius_OHANA_fig1}
\end{center}
\end{figure}

Visual inspection of the Be star observations shows them to be {compatible
  with the canonical picture}, namely a cicumstellar decretion disk. The
targets span all inclinations (equatorial to pole-on) and spectral
subtypes. For some of the brighter stars the disk is already well resolved in
the intermediate configuration (typical baseline lengths 30--70m), and overly
resolved in the large configuration (typical baseline lengths 80--130m) Data
for {$\beta^1$\,Mon and $\mu$\,Cen are shown in
  Fig.~\ref{Rivinius_OHANA_fig1}}.  $\mu$\,Cen shows a broad shallow ramp-type
wing in the line, which is reflected in the phase. This may be the signature
of freshly ejected material closer to the star than the bulk of the disk.

\subsection{OB Supergiants}

The OBA supergiants can be divided into three distinct sets, namely LBVs and
the remaining ones along the wind bi-stability at about $T_{\rm
  eff}=22\,000$\,K.  For the LBVs $\eta$\,Car and HR\,Car, dedicated
contributions to this volume are presented by Mehner et al.\ and Rivinius et
al.  The {O- and early B-supergiants} do not show any obvious signature in
visibility or phase, meaning their winds are too small to detect.  In turn,
the {later B- and A-type supergiants} do show such signatures in visibility,
but again very little in phase. This is obvious for the hypergiant
$\zeta^1$Sco, indicating an extended, but largely symmetric wind
(Fig.~\ref{Rivinius_OHANA_fig2}, left). For HD\,53\,138 this is less obvious
in Fig.~\ref{Rivinius_OHANA_fig2}, right, but comparison with calibrator data
shows the visibility signature to be real, not instrumental. For HD\,53\,138
spectral emission variability goes together with a changing visibility and
phase signature, which may indicate a variable, asymmetric wind.  Further
analysis of the data at hand may provide {constraints on the size and clumping
  of these winds}, while further observations may be able to {trace
  variability}, in particular for the slower winds with flow times of up to
several weeks.

\subsection{Interacting Binaries}
The only observed interacting binary was SS\,Lep, with barely any
interferometric signature. Small wiggles seen in the visibility of some longer
baselines need to be verified.

\section{Conclusions}
The OHANA survey provided interferometric data of the circumstellar
environments of Be stars and OBA supergiants. The raw data is publicly
available, the reduced data will become so as soon as the final reduction has
passed quality control tests. The reduced data will be made available from
{\tt http://activebstars.iag.usp.br/index.php/34-ohana}.

\bibliographystyle{iau307} 
\bibliography{OHANA}

\end{document}